\documentclass{article}
\usepackage[pagewise]{lineno}\nolinenumbers
\usepackage[section]{placeins}
\usepackage{amsmath,amsfonts,amssymb}
  \usepackage{paralist}
  \usepackage{graphics} 
  \usepackage{epsfig} 
\usepackage{graphicx}
\usepackage{subfig}
  \usepackage{epstopdf}
 \usepackage[colorlinks=true]{hyperref}
\hypersetup{urlcolor=blue, citecolor=red}

\newtheorem{theorem}{Theorem}[section]

\newtheorem{proposition}{Proposition}

\newtheorem{remark}{Remark}

\newcommand{\ep}{\varepsilon}
\newcommand{\abs}[1]{\left|#1\right|}

\newcommand{\quot}[1]{``#1''}

\title{Infection model for analyzing biological control of coffee rust using bacterial anti-fungal compounds}

\author{}
\date{}

\begin{document}
\maketitle

\centerline{\scshape Jorge Arroyo Esquivel\footnote{Corresponding author. \url{jorge.arroyoesquivel@ucr.ac.cr}}\footnote{Present address: Universidad de Costa Rica. San Pedro de Montes de Oca, San Jos\'e, Costa Rica, 11501.}}
\medskip
{\footnotesize
 \centerline{Universidad de Costa Rica}
   \centerline{Escuela de Matem\'atica}
   \centerline{San Pedro, San Jos\'e, Costa Rica}
} 

\medskip

\centerline{\scshape Fabio Sanchez}
\medskip
{\footnotesize
 \centerline{Universidad de Costa Rica}
   \centerline{Escuela de Matem\'atica}
   \centerline{San Pedro, San Jos\'e, Costa Rica}
}

\medskip

\centerline{\scshape Luis A. Barboza}
\medskip
{\footnotesize
 \centerline{Universidad de Costa Rica}
   \centerline{Escuela de Matem\'atica}
   \centerline{San Pedro, San Jos\'e, Costa Rica}
}

\bigskip


\begin{abstract}
Coffee rust is one of the main diseases that affect coffee plantations worldwide \cite{cressey2013}. This causes an important economic impact in the coffee production industry in countries where coffee is an important part of the economy. A common method for combating this disease is using copper hydroxide as a fungicide, which can have damaging effects both on the coffee tree and on human health \cite{haddadf2013}. A novel method for biological control of coffee rust using bacteria has been proven to be an effective alternative to copper hydroxide fungicides as anti-fungal compounds \cite{haddad2009}. In this paper, we develop and explore a spatial stochastic model for this interaction in a coffee plantation. We analyze equilibria for specific control strategies, as well as compute the \textit{basic reproductive number}, $\mathcal{R}_0$, of individual coffee trees, conditions for local and global stability under specific conditions, parameter estimation of key parameters, as well as sensitivity analysis, and numerical experiments under local and global control strategies for key scenarios.
\end{abstract}


\section{Introduction}
Throughout the last two centuries, coffee (\textit{Coffea arabica}) has become one of the most extensively produced commodities worldwide, consequently it has helped several developing countries to enter the global market \cite{little1993, pendergrast2010}. During that time, coffee production has been affected by a complex network of diseases and pests \cite{collinge2016, vandermeerj2010b}, from which the coffee rust, a leaf disease caused by the infection of the fungi \textit{Hemileia vastatrix}, has become a major nuisance in the coffee production worldwide, causing serious defoliation in coffee trees, and as a result it causes losses in the crop yield \cite{cressey2013}.

A typical method used for prevention is using fungicides based on copper compounds \cite{haddad2009, loland2004}. The use of these fungicides brings potential problems to the ecosystem through copper enrichment of the soil and health effects on the high consumption of copper \cite{loland2004}, as well as the increase in the population of other coffee pests such as the green scale (\textit{Coccus viridis}) through the destruction of their parasite, the fungi \textit{Lecanicillium lecanii} \cite{jacksond2012, vandermeerj2014}.

As an alternative method of reducing coffee rust and other fungal pathogens in cash crops, several studies have explored the use of bacteria as a biological control method \cite{dorighello2015,haddad2009,jie2013,scholl2012}. The strain of \textit{Bacillus thuringiensis} B157 used as a biological control method has shown to be as effective as copper hydroxide fungicides \cite{haddad2009,haddadf2013}. These bacteria control the rust uredospores by blocking their germination tubes with antibiotic compounds, mainly lipopeptides such as iturine, surfactin, and fengicine \cite{haddadf2013}.

Typically, it is enough to use fungicides based on these compounds to control rust on coffee trees. However, these compounds have been shown to be volatile \cite{filho2010}, making them difficult to be effectively used as an independent fungicide. Moreover, considering the interaction between the bacteria and coffee rust in a coffee plantation, it is possible that one can determine the critical size of bacteria population, this in turn will reduce coffee rust on coffee trees.

Several models have been developed to describe the spread of crop diseases in a field \cite{kranz1990, pielaata1998,smith1998} and some others have been developed to describe the dynamics between the biological control agents and their related pathogens \cite{chatto2002,mills1996,rafikov2008}. However, these models do not propose an interaction between the pathogens and a control agent, in our case the bacteria, while the latter does not consider the spatial distribution of the interacting populations. In this article we develop a stochastic model that considers the interactions between bacteria and coffee rust populations and consider the effect of the spatial distribution of the disease and its spread on the overall dynamics of the disease.

The rest of the paper is organized as follows. In Section 2, we describe the dynamics between the coffee tree, coffee rust, and bacteria. In Section 3, we develop a stochastic epidemic model that illustrates the spread of coffee rust through a coffee plantation, as well as the interaction between the bacteria and coffee rust. Section 4, shows both analytic results and numerical experiments, including theoretical stability analysis for our model, simulations of the coffee rust and bacteria populations, and parameter estimation with its corresponding uncertainty analysis. Lastly, we discuss model results and their biological significance in Section 5.

\section{Coffee rust dynamics}
\subsection{Coffee}

Coffee (\textit{Coffea arabica}) is a perennial tree of the Rubiaceae family originated in the southeast of the Arabic peninsula \cite{clifford1985}. Several types and mutations of this tree have been introduced in several regions of Latin America, Southeast Asia, and Africa since the nineteenth century, from which the berry is used as a food product \cite{pendergrast2010}.

Typical conditions for coffee to be grown depends mostly on altitude and rainfall. Coffee is a highland crop, usually growing between 1000 and 2000 meters of altitude \cite{clifford1985}. In addition, it needs a proper amount of rainfall to grow, where more than four months without rain can cause significant damage to crops \cite{robinson1964}.

\begin{figure}[htp]
\begin{center}
\subfloat[Rainfall distribution of Costa Rica in the province of Alajuela, Costa Rica.]{\includegraphics[width=0.45\textwidth]{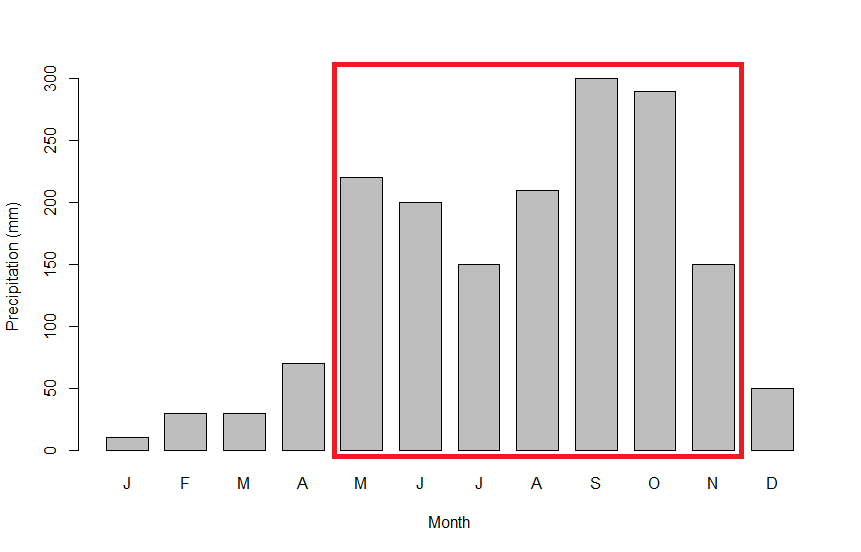}\label{fig:rainparam1}}
	\hspace{0.1cm}
\subfloat[Rainfall distribution of Costa Rica in the province of Cartago, Costa Rica.]{\includegraphics[width=0.45\textwidth]{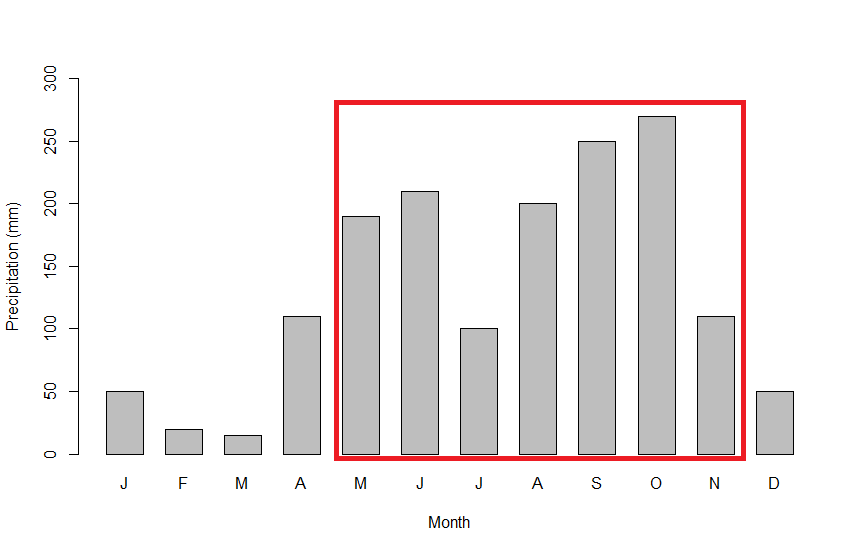}\label{fig:rainparam2}}
\caption{Rainfall distribution of Costa Rica in the provinces of Alajuela and Cartago \cite{imn2017}. These provinces are known for their high coffee production \cite{alvarado2007}. Notice that the months between May and November (inside square) are the months with substantially more rainfall, which correspond to the rainy season in Costa Rica.} \label{fig:rainGraph}
\end{center}
\end{figure}

The yearly rainfall distribution properly defines how the tree produces berries. If there is a single rainfall period in the year, the coffee will only have a single period of harvest, while if there are two periods, it will have two periods \cite{clifford1985}. For example, in the case of Costa Rica, where the rainy season goes from May to October, there is a single harvest season of coffee ranging from June to November, the exact moment depending on the altitude and composition of the soil at which the coffee is being grown \cite{alvarado2007} (see Figure \ref{fig:rainGraph}).

\subsection{Coffee rust}

\begin{figure}[htp]
\begin{center}
\includegraphics[scale=0.8]{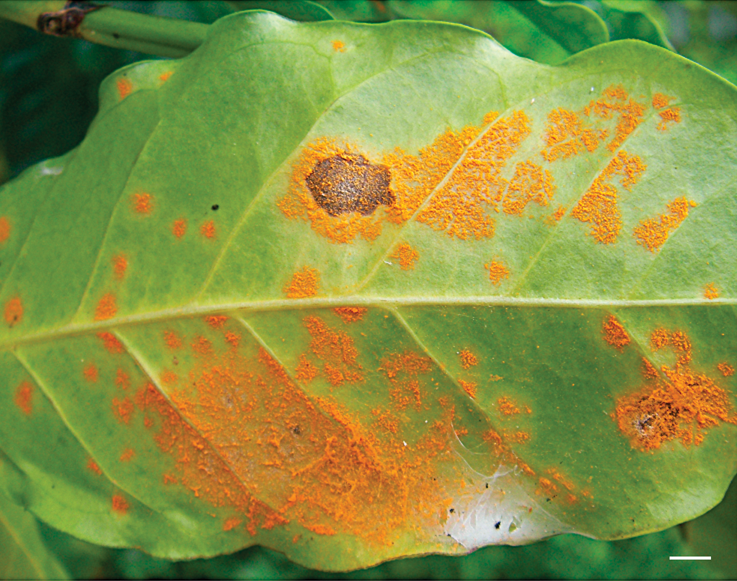}
\caption{Coffee tree leaf infected with an advanced stage of coffee rust \cite{carvalho2011}. This advanced stage shows an intensified color, which gives yellow spots an orange tone.} \label{rustPhoto}
\end{center}
\end{figure}

Coffee trees are affected by several diseases, which may cause significant yield losses in coffee plantations. One of these is the coffee rust, caused by the fungus \textit{Hemileia vastatrix}, which is an obligate parasite that uses coffee as its main host. It was first recorded in 1869 in Ceylon (currently Sri Lanka), where an outbreak suddenly reduced the production of coffee in the island by around 90\% of the total production, causing the coffee industry to cease and made them cultivate tea instead, which is not as profitable \cite{waller1982}. This disease is usually recognizable by yellow spots appearing in the blade of the leaves as shown in Figure \ref{rustPhoto}.

The coffee rust disease is caused by the infection of uredospores of \textit{H. vastatrix} in the leaves. These spores are mainly produced by asexual reproduction of the hyphae \cite{carvalho2011}. For a leaf to be infected, it requires a minimum number of spores to properly perform germination, that number being between 15 and 20 spores \cite{nutmanf1963}. The germination of uredospores depends of several environmental and physiological variables. The main factors influencing the germination process are rainfall density, temperature, and age on both the uredospores and leaves \cite{nutmanf1963}.

For the environmental factors, in low rainfall densities, the germination of uredospores and growth on their mycelium is slower in comparison to the growth in more humid conditions. In cases of extreme dryness, the mycelium of \textit{H. vastatrix} is unable to grow \cite{nutmanf1963}. The optimal temperature for uredospore germination, mycelium growth, and lesion formation in the leaves is 22$^\circ$C. However, with a previous low-temperature stimulation, followed by an increased temperature, the uredospore germination is faster compared to a constant exposition to 22$^\circ$C. This is the case when the plantation receives rainfall during the night, where the temperature is lower, and then warms up during daylight.

In the case of physiological factors, the germination process of uredospores is both slower and have a lower infection rate in the more mature leaves compared to younger leaves \cite{nutmanf1963}. The age of the uredospores plays a role, since the effectiveness of germination of uredospores decrease exponentially.

The uredospores are mostly transmitted inside a single plantation through two main mechanisms, rain splash and wind \cite{nutmanf1963, rayner1961}. However, there is a negative correlation between uredospore age and its ease of being transported through the wind. As it becomes easier to transport the uredospores through the wind, the less effective they are for germination. Hence, we can argue that the principal method for local dispersal is rainfall.

In Costa Rica, an epidemic of coffee rust occurred between 2008 and 2013. During this period, coffee production and price decreased by 16\% and 55\%, respectively. This had an important economic impact on small producers and coffee gatherers \cite{avelino2015}.

The typical method for controlling coffee rust in coffee plantations is by spraying the plantation with several fungicides. In Costa Rica, the fungicide is normally applied at least twice a year, once at the beginning of the rainy season around May, and during September, in the middle of the rainy season as a precautionary measure \cite{icafe2011}. For stricter control measures, more applications can be done.

\subsection{Interaction between bacteria and coffee rust}

The use of bacteria as a biological control method can lead to the reduction of coffee rust in coffee plantations. It has been shown that biological control through the bacteria \textit{Bacillus thuringiensis} results in a process as effective as the application of fungicides of copper hydroxide compounds, which are the standard method of control of coffee rust and has the potential to be harmful to humans \cite{haddad2009}.

This interaction is produced by the effect of several antibiotic compounds produced by the bacteria. In the case of \textit{B. thuringiensis}, these compounds are mostly iturine, surfactine, and fengicine \cite{haddadf2013}. They interact with the germination pores of the uredospores of \textit{H. vastatrix}, blocking the germination tubes inside the uredospore, and therefore stopping the growth of the mycelium.

In the following section we develop a stochastic model that describes this process, allowing us to make predictions about more effective control measures of coffee rust in coffee plantations.

\section{Model}
In this section we introduce a nonlinear differential equation model that describes the spread of coffee rust through a coffee plantation and its interaction with the bacteria \textit{B. thuringiensis}. Previous models have studied a similar phenomenon \cite{chatto2002,mills1996, vandermeer2010b}. Our model is moderately based on the two-population model presented in \cite{vandermeer2010b}. The latter is constructed as follows, let $C$ be the consumer population with carrying capacity $K$, which consumes the resource population with an effective rate $a$ and has a mortality rate $m$. Let $R$ be the resource population, which grows at a rate $r$ and is consumed by the consumer population with a conversion rate $c$. Then the interaction between these two populations is presented by the following system of differential equations:

\begin{equation}
    \begin{split}
        \frac{dC}{dt} =& a\left(\frac{K-C}{K}\right)CR-mC,\\
        \frac{dR}{dt} =& rR(1-R)-a\left(\frac{K-C}{K}\right)CR.
    \end{split}
\end{equation}

Within our model we consider the consumer's growth (in our case the bacteria population) is not affected by the population of the source (in our case the coffee rust population), which means that $a=1$ and $R$ does not affect the logistic growth factor of $C$. Moreover, the source's growth is similar to the growth of the consumer (this means that the growth of the source is also logistic). We also consider the spatial distribution of the coffee rust (source). 

We consider a coffee plantation of $n\times m$ coffee trees, distributed in a rectangular array, where the tree in the $i$-th row and $j$-th column occupies the position $\{i,j\}$. We denote by $H_{\{i,j\}}$ and $B_{\{i,j\}}$ the size of the populations of coffee rust spores \textit{H. vastatrix} and the bacteria \textit{B. thuringiensis} B157 on tree $\{i,j\}$, respectively.

For each $i\in\{1,\ldots,n\},j\in\{1,\ldots,m\}$, both $H_{\{i,j\}}$ and $B_{\{i,j\}}$ are assumed to grow logistically, where $b$ and $h$ are the growth rates and $K_B$ and $K_H$ are the carrying capacities of the bacteria and coffee rust, respectively. Given that coffee rust appears primarily during rainy season, we will consider the carrying capacities constant. We assume that $B_{\{i,j\}}$ has a natural death rate $d$ and the bacteria limits the growth of coffee rust with a constant conversion rate $\gamma$.

Because of the evidence presented in Section 2.2, in our model, the dispersal of coffee rust through wind is neglected. Therefore, the principal method of dispersal is through rainfall, which arrives to the neighbor trees at rate $\alpha$ and is removed from their original tree by rainfall at rate $\beta$. We suppose that the bacteria are being irrigated into the plantation at rate $\mu_{\{i,j\}}$.

The model is then described by the system of nonlinear differential of equations:
\begin{equation}\label{model}
 \begin{split}
\frac{dB_{\{i,j\}}}{dt} =& bB_{\{i,j\}}\left(1-\frac{B_{\{i,j\}}}{K_B}\right)+\mu_{\{i,j\}}(t)-dB_{\{i,j\}},\\
\frac{dH_{\{i,j\}}}{dt} =& hH_{\{i,j\}}\left(1-\frac{H_{\{i,j\}}}{K_H}\right)+\alpha I_{\{i,j\}}-(\gamma B_{\{i,j\}}+\beta) H_{\{i,j\}}.
 \end{split}
\end{equation}

The $I_{\{i,j\}}$ function is defined as a random variable dependent of the coffee tree's neighbors, which correspond to the trees in the cardinal and intermediate directions of the $\{i,j\}$-th tree. This function is the number of spores that can go from any neighbor coffee tree to the $\{i,j\}$-th tree. $I_{\{i,j\}}$ is then defined as:
\begin{equation}\label{immigration}
I_{\{i,j\}} = \sum_{k=-1}^1\sum_{l=-1}^1 H_{\{i+k,j+l\}}(t)\cdot\tau_{\{i+k,j+l\}}(k,l,t)-H_{\{i,j\}}(t)\cdot\tau_{\{i,j\}}(0,0,t),
\end{equation}

\noindent where $H_{\{i,j\}}=0$ if $i\notin\{1,\ldots,n\}, j\notin\{1,\ldots,m\}$. Here, $\tau_{\{ij\}}(k,l,t)$ is a random variable that represents the proportion of coffee rust spores $H_{\{i,j\}}$ that is passed from tree $\{i,j\}$ to tree $\{i+k,j+l\}$.

To define the value of $\tau_{\{i,j\}}$, we will assume that the only method of local dispersal of coffee rust is through rain splash. Therefore, we will base the probability of the random variables $\tau_{\{i,j\}}$ on the dispersal of plant pathogens by the rain splash model provided in \cite{pielaata1998}. In the model, spores splash from a point source and are displaced over a flat ground surface. Each spore is suspended in a water layer, which can be hit by a water splash with probability $\lambda$ and the spore can be lost in the process and stuck on the ground with probability $\ep$.

If the initial distribution of the spores is $\delta(x)$, then the probability per unit length of finding a spore at position $x$ at time $t$ is:
\begin{equation}\label{pielaatProb}
P(t,x) = e^{-\lambda t}\delta(x)+\sum_{i=1}^\infty\frac{(\lambda t)^ie^{-\lambda t}}{i!}\ep^iD(x)^{\star i},
\end{equation}

\noindent where $D(x)^{\star i}$ is the normal distribution convoluted by itself $i$ times, and the normal distribution is given by:
\begin{equation}\label{normalDist}
D(x) = \frac{1}{\sqrt{2\pi\sigma^2}}\exp\left(\frac{-\|x\|^2}{2\sigma^2}\right).
\end{equation}

The model describes how at a point in space $x$, the probability of the spores that were initially at point $x$ that remain at that point decays in an exponential manner and is compensated by the spores from other points that reach this point through rain splash following a Poisson process.

We make additional assumptions to the model. First, we make a discrete version of the model, considering only the spread from a tree to its neighbors, separated by a distance $\|(k,l)\|_{\{i,j\}}$, where $\|\cdot\|_{\{i,j\}}$ is a distance function, not necessarily a norm function.

\begin{remark}
Since the distance between two trees is smaller if they are in the same row, independent of the column, we have that 
$$\|(k,0)\|_{\{i,j\}}\leq \|(0,l)\|_{\{i,j\}}\leq \|(k,l)\|_{\{i,j\}}$$
for $k,l\in\{\pm 1\}$.
\end{remark}

Additionally, if a rust spore is in one tree at time $t$, we assume that in time $t+\Delta t$ there will be at most one splash, where $\Delta t$ is a small change in the time scale (for example, if $t$ is scaled in hours, $\Delta t$ can correspond to a minute). Furthermore, since we only analyze the spread of coffee rust in time $t+\Delta t$, we do not consider the exponential decay value, which begins at time $0$. Hence, we only consider the case $i = 1$ in Equation \ref{pielaatProb}.

In the case of no rain, the dispersal is negligible. To consider the chance that there is no rain, assume there is a probability $P_\lambda(t)$ at time $t$ of raining. Given these assumptions, our random variables $\tau_{\{i,j\}}$ satisfy that
$$\tau_{\{i,j\}}(k,l,t)\sim\text{Bernoulli}(P_{\{i,j\}}(k,l,t)),$$
where:
\begin{equation}\label{dispersalProb}
P_{\{i,j\}}(k,l,t) = \psi\lambda(t)H_{\{i,j\}}(t)D\left(\|(k,l)\|_{\{i,j\}}\right),
\end{equation}

\noindent and $\psi$ is a proportionality constant, $\lambda(t)\sim\text{Bernoulli}(P_\lambda(t))$ is the probability that it is raining at time $t$, and $D$ is the normal distribution with mean $0$ and variance $\sigma^2$ as in Equation \ref{normalDist}. Note that $P_\lambda$ depends on time, which is important since rainfall amount is dependent on the period of the year, as shown in Figure \ref{fig:rainGraph}.

For the values of $\mu_{\{i,j\}}$, each value will define a specific control strategy. First, we define a global control strategy where the whole plantation is irrigated at the same rate, this means, for all $\{i,j\}$:
\begin{equation}\label{globalControl}
\mu_{\{i,j\}}\equiv \mu
\end{equation}

\noindent for some constant $\mu$. This strategy works as a preemptive measure, preparing the coffee trees to counter coffee rust. Then the local control strategy, which defines a rate that is proportional to the amount of coffee rust in the tree itself and its neighbors, for all $\{i,j\},$ is given by:
\begin{equation}\label{localControl}
\mu_{\{i,j\}}(t) = \rho H_{\{i,j\}}(t)+\delta\left(\sum_{k=-1}^1\sum_{l=-1}^1 H_{\{i+k,j+l\}}(t)-H_{\{i,j\}}(t)\right),
\end{equation}

\noindent for some constants $\rho,\delta$ defined as the intensity of spraying on a tree and its neighbors, respectively. This control strategy works as a reactive strategy, where the trees are irrigated depending on its neighbors’ infectious status.

\section{Results}
In this section we refer to the mathematical and numerical results of our study. First, we show the theoretical equilibria of the model under necessary conditions using the \textit{basic reproductive number} and prove global stability of the disease-free equilibrium, as well as the local stability of the endemic equilibrium in a specific plantation size. Moreover, we perform several numerical experiments under local and global strategies for the control of coffee rust on coffee plantations to study the temporal and spatial dynamics of our model, as well as parameter estimation of the model under certain conditions to analyze the sensitivity of the parameters in the overall behavior of the system.

\subsection{Bacteria and coffee rust equilibrium}
Given the recursive nature of the model, we only study the following case. Consider a global control strategy as shown in Equation \ref{globalControl}, and take $P_\lambda\equiv 1$ as the probability for $\lambda$ in Equation \ref{dispersalProb}, which implies the plantation has a constant amount of rainfall. More so, we assume that the difference between the distances of neighbor trees from a specific tree are negligible, this means $\left|D(\|(k_1,l_1)\|_{\{i,j\}})-D(\|(k_2,l_2)\|_{\{i,j\}})\right|<\ep$ for small enough $\ep$ and $k_i,l_i\in\{\pm1,0\}$.

Since $\mu_{\{i,j\}}$ is constant, we get that:
$$\frac{dB_{\{i,j\}}}{dt} = \frac{dB_{1,1}}{dt}$$

\noindent for all $i\in\{1,\ldots,n\},j\in\{1,\ldots,m\}$. Then, for all $i,j$, if $b>d$, the equilibrium point of the bacteria population is the following:
\begin{equation}\label{Bequilibrium}
B_{\{i,j\}}^* = B = \frac{1}{2b}K_B\left(b-d\right)\left(1+\sqrt{1+\frac{4b\mu}{K_B(b-d)^2}}\right).
\end{equation}

We only consider positive equilibria; hence it is clear from Equation \ref{Bequilibrium} that this equilibrium exists whenever the birth rate of bacteria is bigger than its death rate.
\begin{proposition}
The bacteria equilibrium shown in Equation \ref{Bequilibrium} exists if and only if $b> d$. 
\end{proposition}
For the coffee rust population, since $P_\lambda\equiv 1$, and the difference in tree distances $D(\|(k,l)\|_{\{i,j\}})$, are negligible, the probability of dispersal is homogeneous throughout the coffee plantation. Therefore, we can conclude that in equilibrium the value of $I_{\{i,j\}}$ behaves in a similar manner to a constant $\Pi$ for all $i,j$, where $\Pi$ corresponds to the proportion of coffee rust spores each tree has if the only dynamic mechanism considered is dispersal (this means, growth and competition between populations is not considered). Hence, the equilibria of each coffee rust population is given by:
\begin{equation}\label{Hequilibrium}
H_{\{i,j\}}^* = \frac{K_H}{\gamma Bh}(h+\Pi\alpha-\beta-\gamma B).
\end{equation}

Define $F_0 = h+\Pi\alpha-\beta$ as the net growth rate of the coffee rust spore population on each tree. Then, if $B>0$ (this means, there is bacteria in equilibrium), the \textit{basic reproductive number} of coffee rust is defined as the average number of infected trees the population of coffee rust in a single tree can produce and is given by:
\begin{equation}\label{R0}
R_0 = \frac{F_0}{\gamma B},
\end{equation}

which turns Equation \ref{Hequilibrium} into:
\begin{equation}\label{Hequilibrium2}
H_{\{i,j\}}^* = \frac{K_H}{h}(R_0-1).
\end{equation}

From this equation we get the following theorem.
\begin{theorem}
Given $R_0$ as in Equation \ref{R0}, then the endemic coffee rust equilibrium exists if and only if $R_0>1$.
\end{theorem}

Note that, by taking $H_{\{i,j\}}\equiv0$ for all $\{i,j\}$, then it is shown that the disease-free equilibrium always exists.

\subsection{Global stability of disease-free equilibrium}
In this section we prove that, for any plantation, the disease-free equilibrium is globally asymptotically stable. To do this, we use the method described on the main result in \cite{ccc2002}:

\begin{theorem}
Let $X\in\mathbb{R}^n$ be the uninfected individuals and $Z\in\mathbb{R}^m$ the infected individuals in the system such that the System \ref{model} is rewritten as:

\begin{equation}\label{ccc}
\begin{split}
    \frac{dX}{dt} =& F(X,Z),\\
    \frac{dZ}{dt} =& G(X,Z).
\end{split}
\end{equation}

Then, if the three following conditions are met:

\begin{enumerate}
    \item $\mathcal{R}_0<1$
    \item For $\frac{dX}{dt} = F(X,0)$, the disease-free equilibrium $X^*$ is globally asymptotically stable.
    \item $G(X,Z) = AZ-\hat{G}(X,Z)$, where $A = G_Z(X^*,0)$ and $\hat{G}(X,Z)\geq 0$ for all $(X,Z)$ where the model makes sense.
\end{enumerate}

Then the disease-free equilibrium is globally asymptotically stable.
\end{theorem}

In this section we will prove that conditions (2) and (3) are met by our model. To do this, starting from the $\{1,1\}$-th tree, relabel the trees from the first column down to the $\{n,1\}$-th tree as $1,\ldots,n$. Then, starting from the $\{1,2\}$-th tree relabel the trees from the second column down to the $\{n,2\}$-th tree as $n+1,\ldots,2n$ and repeat this process up to the $m$-th column. Then we write $X=(B_1,\ldots,B_{mn})$ and $Z=(H_1,\ldots,H_{mn})$.

To prove (2), note that for all $i$, by repeatedly derivating we get that

$$\frac{d^kB_i}{dt^k}=0$$ 

\noindent in the equilibrium (when $B_i=B^*$ with $B^*$ as in Equation \ref{Bequilibrium}). Hence, we have that in the equilibrium, $B_i$ becomes constant. Therefore $B_i\rightarrow{B^*}$ when $t\rightarrow{\infty}$ and the equilibrium $X^*$ is globally asymptotically stable.

To prove (3), given that we consider only the asymptotic behavior, we can consider that $I_{\{i,j\}} =\Pi$ and $B_i= B_i^*$. Note that for all $i$, the $i$-th entry of the vector $G$ equals

$$G(X,Z)_i=hH_i\left(1-\frac{H_i}{K_H}\right)+\alpha\Pi\sum_{\text{ neighbor}_i}H_j-(\beta+B_i)H_i,$$

where $j\in\text{neighbor}_i$ if the following condition is met:

$$\begin{cases}
\abs{i-j}\in\{1,n-1,n+1\}\hbox{ if }i\not\equiv0,1\mod n\\
\abs{i-j}\in\{n-1,n+1\}\hbox{ or }i-j=-1\hbox{ if }i\equiv1\mod n\\
\abs{i-j}\in\{n-1,n+1\}\hbox{ or }i-j=1\hbox{ if }i\equiv0\mod n
\end{cases}.$$

Then, it follows that:

$$(A)_{i,j} =\begin{cases}
h-(\beta+B_i^*)\hbox{ if }i=j\\
\alpha\Pi\hbox{ if }j\in\text{neighbor}_i\\
0\hbox{ if else}
\end{cases}.$$

Therefore

$$(AZ)_i = hH_i+\alpha\Pi+\sum_{\text{ neighbor}_i}H_j-(\beta+B_i^*)H_i.$$

Take

$$\hat{G}(X,Z)_i = \frac{hH_i}{K_H}$$

Therefore, we have that the system satisfies condition (3). This lets us conclude the following theorem:

\begin{theorem}
If $\mathcal{R}_0<1$, the disease-free equilibrium is globally asymptotically stable.
\end{theorem}

\subsection{Local stability of endemic equilibrium in a $1\times 2$ system}
The structure of the model gives an additional difficulty when analyzing the stability of the endemic equilibrium when $\mathcal{R}_0>1$. However, the following result can be derived in a $1\times2$ system.

\begin{theorem}
If $(n,m)\in\{(1,2),(2,1)\}$ and $\mathcal{R}_0>1$, then if

$$1<\frac{2}{\gamma B^*}$$

\noindent then the endemic equilibrium is locally asymptotically stable. Otherwise, if

$$1>\frac{2}{\gamma B^*}$$

\noindent then the endemic equilibrium is unstable.
\end{theorem}

To prove this result, we will consider that $B_i=B^*$ for $i\in\{(1,2)\}$. Then, the linearization of the $H_i$ values comes as:

$$J=\left(\begin{array}{cc}
h\left(1-\frac{2H^*}{K_H}\right)-\beta-\gamma B^* & \alpha\Pi\\
\alpha\Pi & h\left(1-\frac{2H^*}{K_H}\right)-\beta-\gamma B^*
\end{array}\right).$$

This matrix has eigenvalues

$$h\pm\alpha\Pi-\left(\beta+\gamma B^*+\frac{2hH^*}{K_H}\right).$$

Since all parameters are positive, notice that the biggest eigenvalue is found when $\pm\alpha\Pi$ is positive. Notice that this eigenvalue is negative when

$$h+\alpha\Pi<\beta+\gamma B^*+\frac{2hH^*}{K_H}.$$

\noindent If we order the terms to get $\mathcal{R}_0$ as in Equation \ref{R0} we get that:

$$\mathcal{R}_0<1+\frac{2hH^*}{\gamma K_HB^*}.$$

By making the substitution of $H^*$ as in Equation \ref{Hequilibrium2}, we get that:

$$\mathcal{R}_0\left(1-\frac{2}{\gamma B^*}\right)<1-\frac{2}{\gamma B^*}.$$

Which is equivalent to $\mathcal{R}_0>1$ if $1-\frac{2}{\gamma B^*}<0$. Otherwise if $1-\frac{2}{\gamma B^*}>0$, then we get that $\mathcal{R}_0<1$, which is a contradiction, therefore both eigenvalues are positive, and the endemic equilibrium is unstable.

\subsection{Numerical simulations}
\begin{figure}[htp]
\begin{center}
\subfloat{\includegraphics[width=0.4\textwidth]{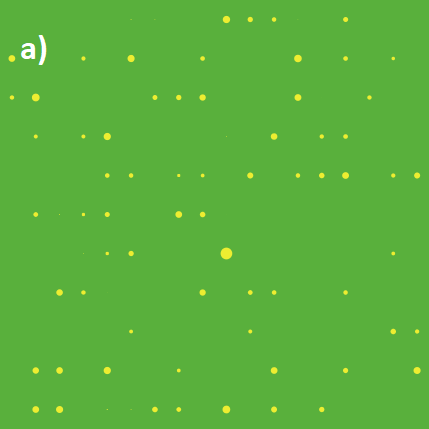}\label{fig:initialglobalplantation}}
  \quad
\subfloat{\includegraphics[width=0.4\textwidth]{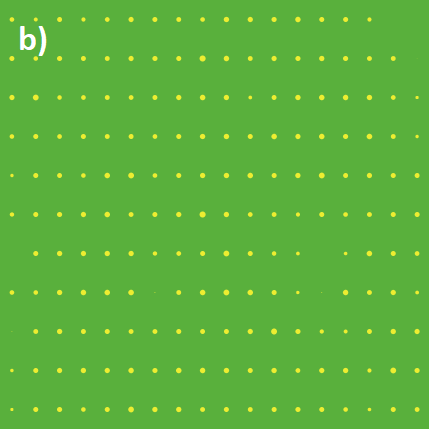}\label{fig:advancedglobalplantation}}\\
\subfloat{\includegraphics[width=0.4\textwidth]{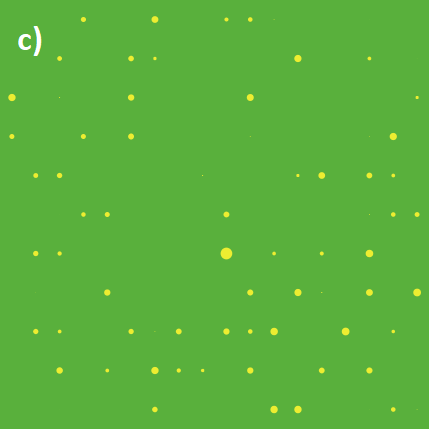}\label{fig:initialplantation}}
  \quad
\subfloat{\includegraphics[width=0.4\textwidth]{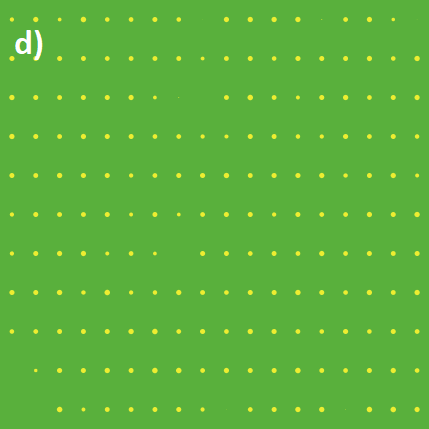}\label{fig:advancedplantation}}
\caption{Examples of a $11\times 18$ plantation simulated in NetLogo. Figures \ref{fig:initialglobalplantation} and \ref{fig:initialplantation} represent the initial conditions (with 50\% of the plantation infected with coffee rust) of the plantation in global and local control, respectively, and Figures \ref{fig:advancedglobalplantation} and \ref{fig:advancedplantation} represent the simulated plantations after 72 hours using global and local control, respectively. The trees are uniformly distributed in the plantation, each yellow point represents an infected tree. The size of the yellow point represents the proportion of coffee rust in each tree.} \label{fig:plantExample}
\end{center}
\end{figure}

Numerical experiments are performed using NetLogo \cite{netlogo}. In each simulation, for simplicity of computation and visualization we will consider a $11\times 18$ plantation, which corresponds to approximately $4\%$ of the recommended amount of coffee trees in a hectare \cite{icafe2011}. The simulation display in Netlogo is shown in Figure \ref{fig:plantExample}. Model parameters are summarized in Table \ref{param}.

{\renewcommand{\arraystretch}{1.3}
\begin{table}[htp]
\begin{center}
\begin{tabular}{|c p{5cm} p{5cm} c|}
\hline
Parameter & Description & Value(s) & Source\\
\hline
$b$ & Natural growth rate of bacteria & $2 \hbox{ hour}^{-1}$& \cite{li2012}\\
$K_B$ & Carrying capacity of bacteria & $7 \log\hbox{bacteria}$ & \cite{li2012}\\
$d$ & Natural death rate of bacteria & $1.5 \hbox{ hour}^{-1}$ & \cite{li2012}\\
$K_H$ & Carrying capacity of coffee rust & $\{0.2380, 0.2840\} \hbox{ rust proportion}$ & \cite{haddad2009, nutmanf1963}\\
$h$ & Natural growth rate of rust & $(0,0.35) \hbox{ hour}^{-1}$ & \cite{nutmanf1963}\\
$\alpha$ & Immigration rate of coffee rust & $(0,1)\hbox{ day}^{-1}$ & \\
$\beta$ & Emigration rate of coffee rust& $(0,1)\hbox{ day}^{-1}$ & \\
$\psi$ & Proportionality constant of dispersal probability & $[0,20]\hbox{ rust proportion}^{-1}\hbox{ m}$ & \\
$\gamma$ & Conversion rate of bacteria to coffee rust & $(0,1)\hbox{ log bacteria}^{-1}\hbox{ day}^{-1}$ & \\
$\sigma$ & Standard deviation of dispersal probability of rust & $3\hbox{m}^{-1}$ & \cite{pielaata1998}\\
$d_{ij}(0,\pm 1)$ & Distance between columns of coffee trees & $2 \hbox{ m}$ & \cite{icafe2011}\\
$d_{ij}(\pm 1,0)$ & Distance between rows of coffee trees &$1 \hbox{ m}$ & \cite{icafe2011}\\
$d_{ij}(\pm 1,\pm 1)$ & Diagonal distance between coffee trees & $2.2361 \hbox{ m}$ & \cite{icafe2011}\\
\hline
\end{tabular}
\caption{Parameters values.}
\label{param}
\end{center}
\end{table}
}

For all simulations, we will assume that initially $50\%$ of the plantation is infected, which corresponds to a value like that reported in \cite{vandermeerj2014}, with different initial values of coffee rust for each $\{i,j\}$, as shown in Figure \ref{fig:plantExample}. We will analyze both local and global control strategies in different settings (high/low growth rate of coffee rust, high/low dispersal of coffee rust, high/low conversion rate of bacteria).

\begin{figure}[htp]
\begin{center}
\subfloat[Proportion of infected trees versus time (hours)]{\includegraphics[width=0.5\textwidth]{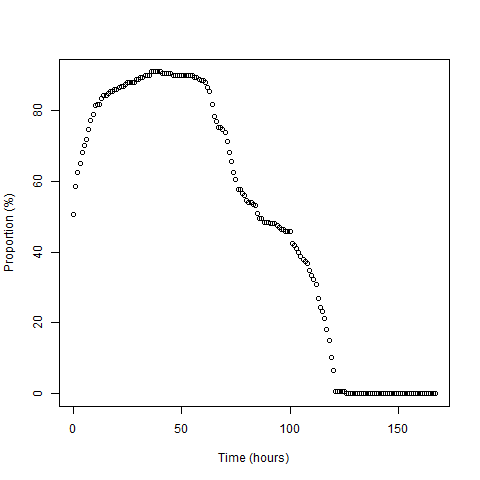}\label{fig:21plants}}
  \hfill
\subfloat[Proportion of leaves covered by rust versus time (hours)]{\includegraphics[width=0.5\textwidth]{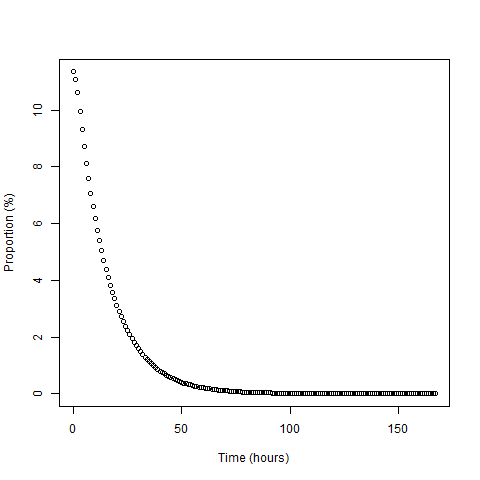}\label{fig:21rust}}
\caption{Time series of simulation using $h = 0.01$, $K_H = 0.238$, $\alpha = 0.5$, $\beta = 0.1$, $\gamma = 0.7$, $\psi = 20$, $\mu = 0.5, P_\lambda\equiv 1$. $\mathcal{R}_0=1.07$ 95\% CI $[1.02,1.13]$.} \label{fig:sim21}
\end{center}
\end{figure}

\begin{figure}[htp]
\begin{center}
\subfloat[Proportion of infected trees versus time (hours)]{\includegraphics[width=0.5\textwidth]{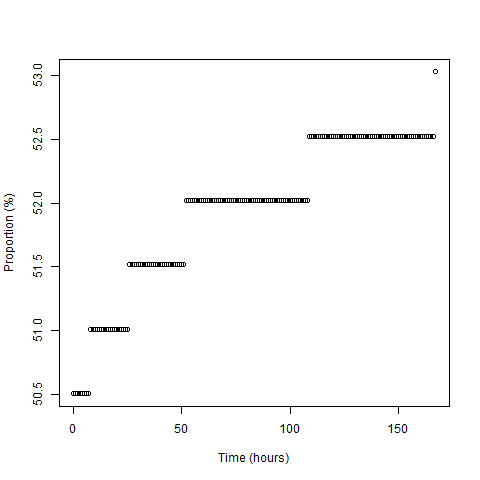}\label{fig:43plants}}
  \hfill
\subfloat[Proportion of leaves covered by rust versus time (hours)]{\includegraphics[width=0.5\textwidth]{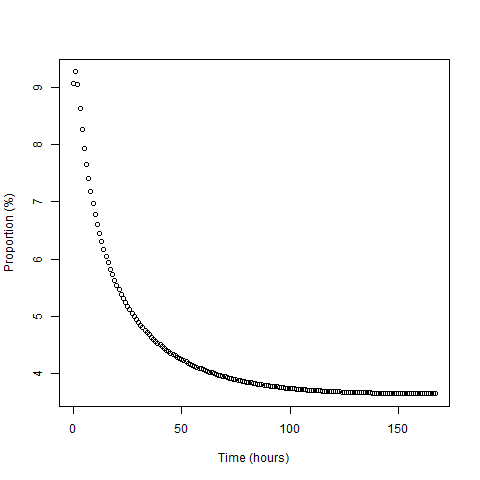}\label{fig:43rust}}
\caption{Time series of simulation using $h = 0.15$, $K_H = 0.284$, $\alpha = 0.9$, $\beta = 0.5$, $\gamma = 0.9$, $\psi = 0.1$, $\mu = 1, P_\lambda\equiv 1$. $\mathcal{R}_0=1.01$ 95\% CI $[1.00,1.03]$.} \label{fig:sim43}
\end{center}
\end{figure}

\begin{figure}[htp]
\begin{center}
\subfloat[Proportion of infected trees versus time (hours)]{\includegraphics[width=0.5\textwidth]{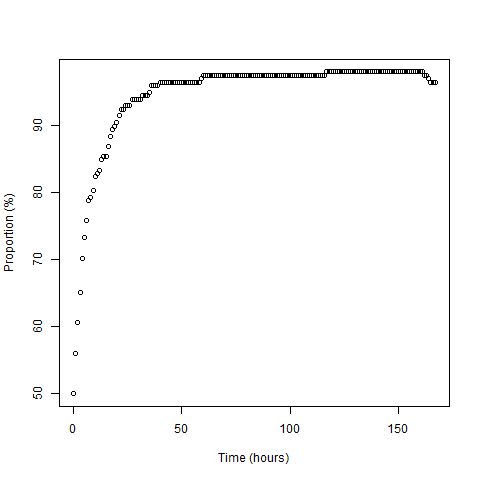}\label{fig:110plants}}
  \hfill
\subfloat[Proportion of leaves covered by rust versus time (hours)]{\includegraphics[width=0.5\textwidth]{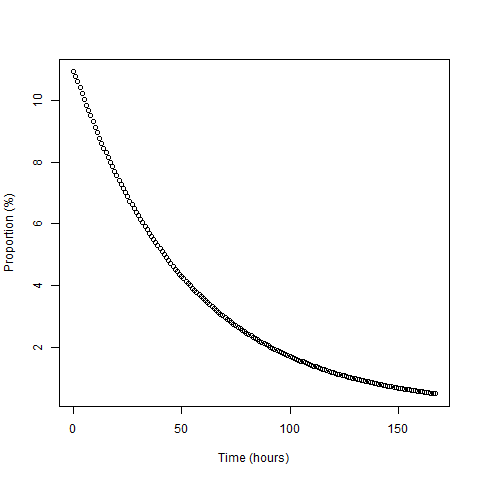}\label{fig:110rust}}
\caption{Time series of simulation using $h = 0.01$, $K_H = 0.238$, $\alpha = 0.5$, $\beta = 0.5$, $\gamma = 0.1$, $\psi = 20$, $\rho = 0.9$, $\delta = 0.2$, $P_\lambda\equiv 1$. $\mathcal{R}_0=1.04$ 95\% CI $[1.00,1.07]$).} \label{fig:sim110}
\end{center}
\end{figure}

\begin{figure}[htp]
\begin{center}
\subfloat[Proportion of infected trees versus time (hours)]{\includegraphics[width=0.5\textwidth]{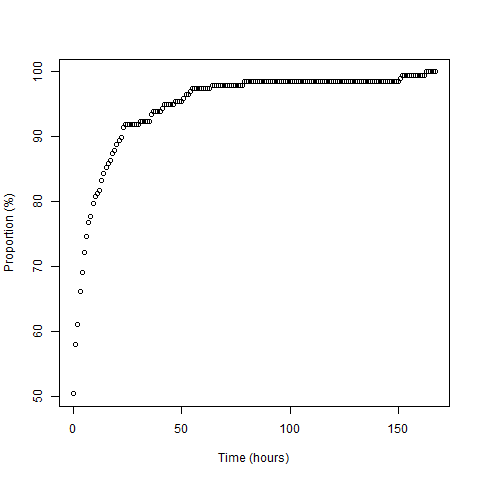}\label{fig:115plants}}
  \hfill
\subfloat[Proportion of leaves covered by rust versus time (hours)]{\includegraphics[width=0.5\textwidth]{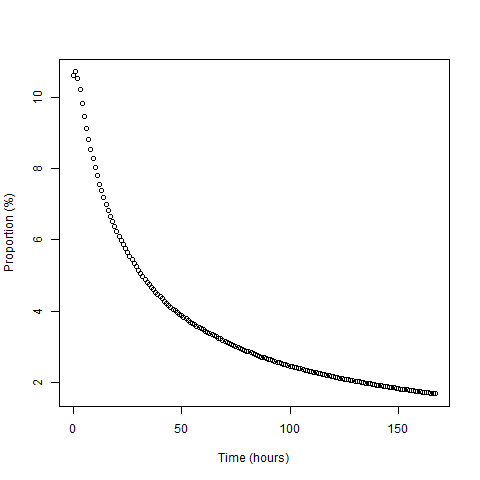}\label{fig:115rust}}
\caption{Time series of simulation using $h = 0.07$, $K_H = 0.284$, $\alpha = 0.9$, $\beta = 0.1$, $\gamma = 0.9$, $\psi = 20$, $\rho = 0.9$, $\delta = 1$, $P_\lambda\equiv 1$. $\mathcal{R}_0=1.04$ 95\% CI $[1.00,1.07]$.} \label{fig:sim115}
\end{center}
\end{figure}

\begin{figure}[htp]
\begin{center}
\subfloat[Proportion of infected trees versus time (hours)]{\includegraphics[width=0.5\textwidth]{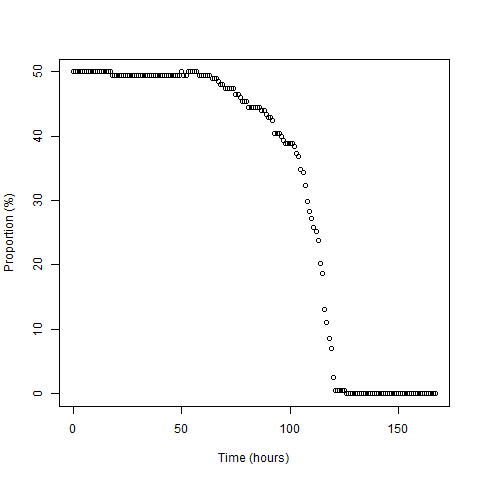}\label{fig:stoc1Plants}}
  \hfill
\subfloat[Proportion of leaves covered by rust versus time (hours)]{\includegraphics[width=0.5\textwidth]{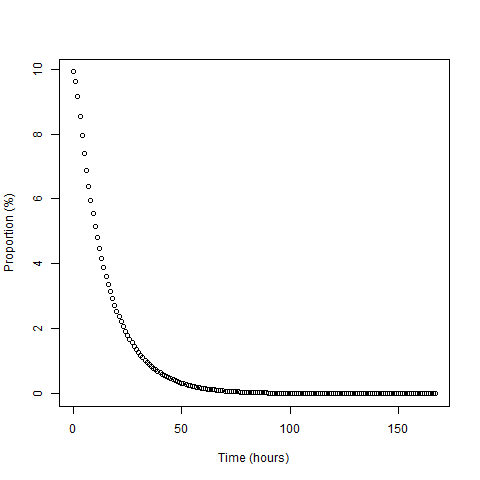}\label{fig:stoc1rust}}
\caption{Time series of simulation using $h = 0.01$, $K_H = 0.238$, $\alpha = 0.5$, $\beta = 0.1$, $\gamma = 0.7$, $\psi = 20$, $\mu = 0.5, P_\lambda\equiv 0.6$. $\mathcal{R}_0=0.98$ 95\% CI $[0.92,1.03]$.} \label{fig:simstoc1}
\end{center}
\end{figure}

\begin{figure}[htp]
\begin{center}
\subfloat[Proportion of infected trees versus time (hours)]{\includegraphics[width=0.5\textwidth]{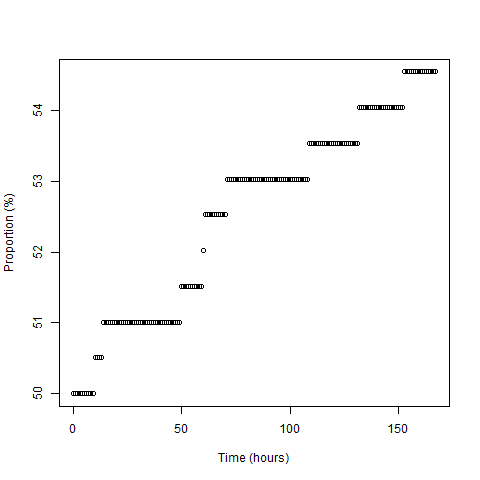}\label{fig:stoc10Plants}}
  \hfill
\subfloat[Proportion of leaves covered by rust versus time (hours)]{\includegraphics[width=0.5\textwidth]{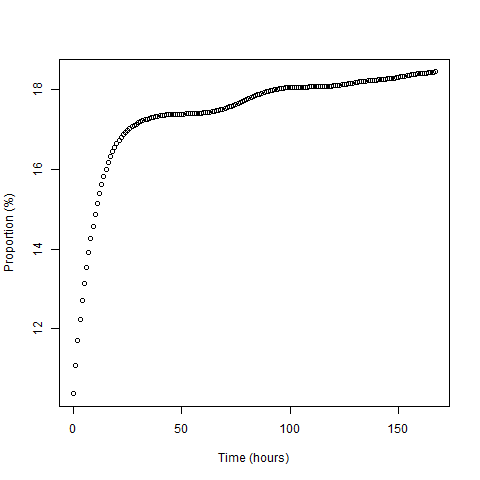}\label{fig:stoc10rust}}
\caption{Time series of simulation using $h = 0.3$, $K_H = 0.238$, $\alpha = 0.5$, $\beta = 0.9$, $\gamma = 0.5$, $\psi = 0.1$, $\rho = 0.8$, $\delta = 0.4, P_\lambda\equiv 0.645$. $\mathcal{R}_0=1.00$ 95\% CI $[0.98,1.03]$.} \label{fig:simstoc10}
\end{center}
\end{figure}

\begin{figure}[htp]
\begin{center}
\subfloat[Proportion of infected trees versus time (hours)]{\includegraphics[width=0.5\textwidth]{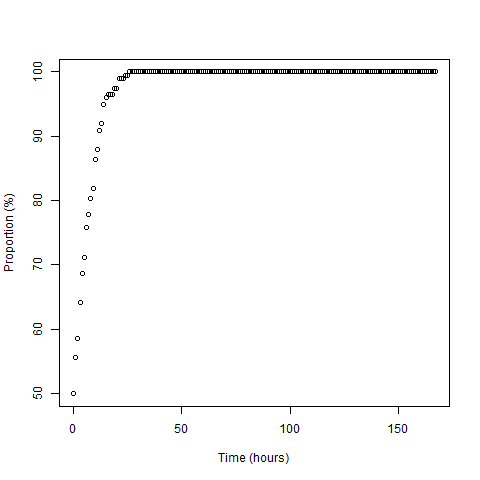}\label{fig:127Plants}}
  \hfill
\subfloat[Proportion of leaves covered by rust versus time (hours)]{\includegraphics[width=0.5\textwidth]{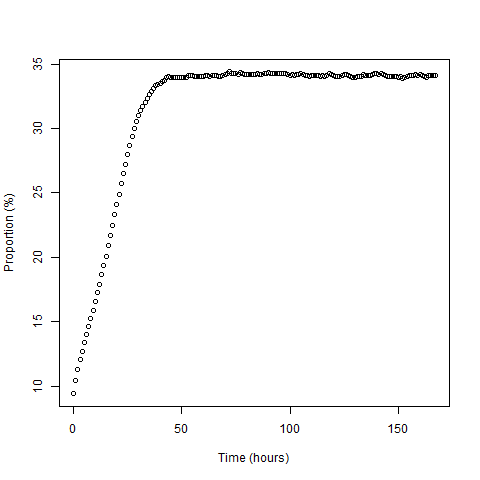}\label{fig:127rust}}
\caption{Time series of simulation using $h = 0.3$, $K_H = 0.238$, $\alpha = 0.9$, $\beta = 0.1$, $\gamma = 0.9$, $\psi = 20$, $\rho = 0.2$, $\delta = 0.7, P_\lambda\equiv 1$. $\mathcal{R}_0=1.06$ 95\% CI $[1.01,1.10]$.} \label{fig:sim127}
\end{center}
\end{figure}

The time scale of the simulations is in hours. We explore the behavior of the system in a small-time interval, hence, the duration of each simulation will be one week (this means $t\in [0,168]$). In each simulation we analyze the proportion of leaves infected by coffee rust (with the assumption that on average all trees have the same amount of leaves) and the proportion of trees being infected by coffee rust. An advanced state of this simulation in Netlogo (this means, after 72 hours of execution) is seen in Figure \ref{fig:advancedglobalplantation}. Figures \ref{fig:sim21}, \ref{fig:sim43}, and \ref{fig:simstoc1} show time series of simulations where global control is applied, while Figures \ref{fig:sim110}, \ref{fig:sim115},  \ref{fig:simstoc10}, and \ref{fig:sim127} show time series of simulations where local control is applied. We also calculate the \textit{basic reproductive number} $\mathcal{R}_0$ on each simulation using the Maximum Likelihood estimation method presented in the R0 package of R \cite{obadia2012}. This method provides a workaround to the difficulty of estimating $\mathcal{R}_0$ in an analytical manner.

Figure \ref{fig:sim21} shows a scenario where the coffee rust growth rate is small, but its dispersal rate and probability of dispersal are high. These values are reflected by how strongly coffee rust is spread throughout the first days of the simulation, but the inhibition produced by bacteria with a high conversion rate gets into a breaking point, where the proportion of infected trees decrease rapidly. This is also accompanied by a slow decrease of coffee rust through the plantation. This suggests that although the dispersal is effective, the amount of coffee rust spores in the original trees decays, and under some threshold value, the coffee rust population is small enough that it is no longer able to support itself in a particular tree, therefore reducing the proportion of infected trees considerably.

In Figure \ref{fig:sim43} we see a different behavior. This is explained by the high coffee rust growth rate, which compensates for the loss obtained by the conversion produced by bacteria, avoiding the coffee rust on each individual tree to pass through a critical point where coffee rust spores can no longer sustain itself on the leaves. Also notice that for small values of $\psi$ (overall probability of dispersal by rain splash) makes the spread of the coffee rust spores behave in stochastic jumps, rather than a continuous function. 

Figure \ref{fig:sim110} showcases how the coffee rust decaying rate is affected primarily by the conversion rate. Here, both coffee rust growth rate and conversion rate are small, hence the main source of coffee rust decay is the emigration rate ($\beta$). Since the values of $\psi$ and immigration rate are both high, the coffee rust spores propagate through the plantation rapidly. Notice that in the end of the simulation the proportion of infected trees starts to go down, possibly as a delayed decay that happened after coffee rust reaches its threshold.

In Figure \ref{fig:sim115}, the coffee rust growth rate is in the middle of the considered range, with a low emigration rate. This appears to be enough to compensate the high conversion rate, which makes the coffee rust decay. Furthermore, the high immigration rate ($\alpha$) and overall probability of dispersal by rain splash ($\psi$) help the coffee rust spores reach most of the plantation.

Figure \ref{fig:simstoc1} shows how the trend of the previous simulations does not appear when there are rainless periods. In this scenario the proportion of infected trees remains relatively constant throughout several days and suddenly decays given the small amount of coffee rust spores. This happens even if the value of $\psi$ is high, which is related to an increased dispersal of coffee rust spores. It is important to note that $P_\lambda<1$ is a more realistic scenario for ranges of time like weeks, where the rainfall is not constant throughout the day.

Figure \ref{fig:simstoc10} shows a scenario where the coffee rust growth rate is high, which avoids the compensation of other values that inhibit the coffee rust's growth, such as emigration rate and conversion rate. Note that in this scenario, there was more rain than in Figure \ref{fig:simstoc1}, which can be seen as having more opportunities of dispersal (hence more jumps in the graph), even though the value of $\psi$ was smaller. In this scenario coffee rust was successful and managed to get through the control of the bacteria, suggesting that with a high growth rate, coffee rust could not be controlled by this method.

In the case of Figure \ref{fig:sim127}, both the coffee rust growth rate and the conversion rate of bacteria are high. In this scenario coffee rust eventually takes over the plantation, and reaches its carrying capacity in two days. This suggests that a high growth rate of coffee rust is not feasible, since no conversion rate from the bacteria seems to be able to compensate the rate at which coffee rust grows on each tree.

Our simulations show that when the plantation is at coffee rust-free equilibrium, that is, $H_{\{i,j\}}=0$ for all $\{i,j\}$, it is indicative that a coffee rust threshold exists where the coffee rust incidence on each tree can sustain itself. This suggests there is a minimum viable population of coffee rust in the coffee plantation.

\begin{figure}[htp]
\begin{center}
\subfloat{\includegraphics[width=0.4\textwidth]{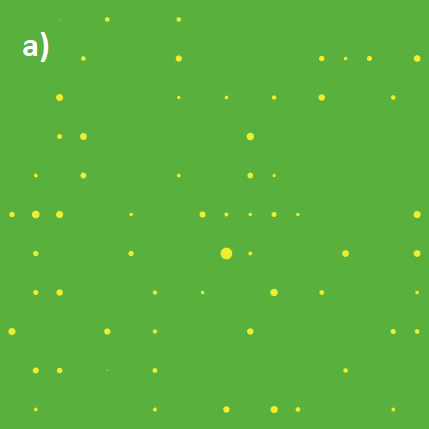}\label{fig:global50initial}}
  \quad
\subfloat{\includegraphics[width=0.4\textwidth]{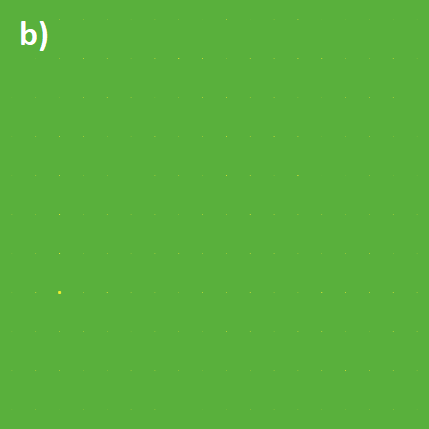}\label{fig:global50final}}\\
\subfloat{\includegraphics[width=0.4\textwidth]{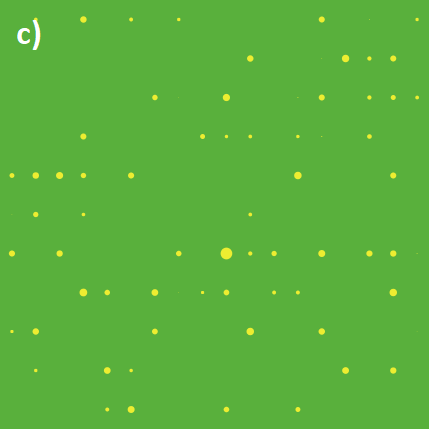}\label{fig:local50initial}}
  \quad
\subfloat{\includegraphics[width=0.4\textwidth]{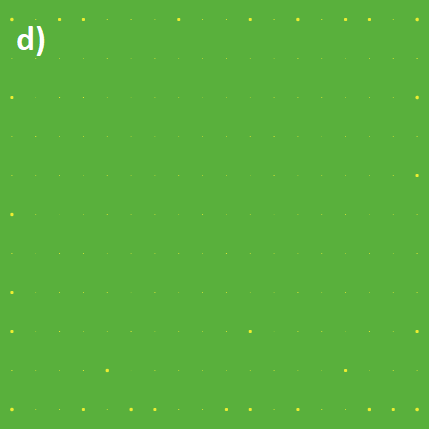}\label{fig:local50final}}
\caption{Simulated plantations with 50\% of the plantation infected with coffee rust using $h = 0.15,K_H = 0.284,\alpha = 0.9,\beta = 0.1,\gamma = 0.5, \psi = 20$, $\mu = 2.5$ for global control, and  $\rho = 7,\delta = 7$ for local control. Figures \ref{fig:global50initial} and \ref{fig:local50initial} show the initial distribution of coffee rust in a plantation with global and local control, respectively, and Figures \ref{fig:global50final} and \ref{fig:local50final} show the same plantations after 168 hours.} \label{fig:plantation50}
\end{center}
\end{figure}

\begin{figure}[htp]
\begin{center}
\subfloat{\includegraphics[width=0.4\textwidth]{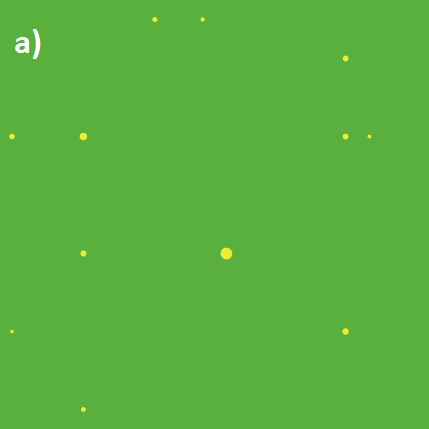}\label{fig:global10initial}}
  \quad
\subfloat{\includegraphics[width=0.4\textwidth]{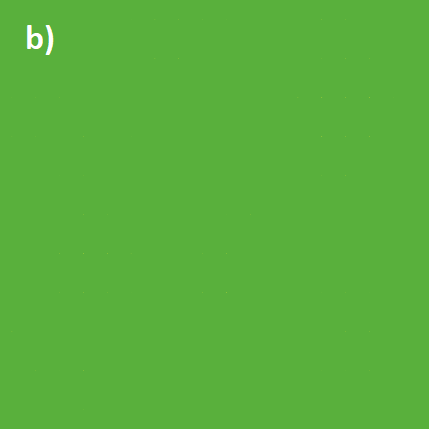}\label{fig:global10final}}\\
\subfloat{\includegraphics[width=0.4\textwidth]{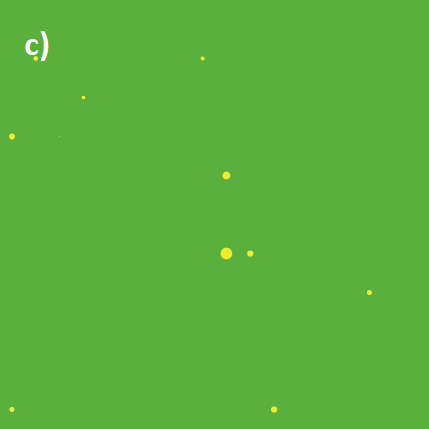}\label{fig:local10initial}}
  \quad
\subfloat{\includegraphics[width=0.4\textwidth]{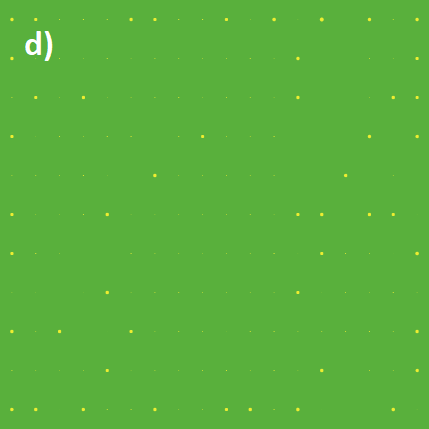}\label{fig:local10final}}
\caption{Simulated plantations with 10\% of the plantation infected with coffee rust using $h = 0.15,K_H = 0.284,\alpha = 0.9,\beta = 0.1,\gamma = 0.5, \psi = 20$, $\mu = 2.5$ for global control, and  $\rho = 7,\delta = 7$ for local control. Figures \ref{fig:global10initial} and \ref{fig:local10initial} show the initial distribution of coffee rust in a plantation with global and local control, respectively, and Figures \ref{fig:global10final} and \ref{fig:local10final} show the same plantations after 168 hours.} \label{fig:plantation10}
\end{center}
\end{figure}

We also performed experiments for high values of $h$ and $\gamma$ under values of the control parameters above our defined framework and with different initial conditions of coffee rust in the plantation. To do this, we let $h = 0.15,K_H = 0.284,\alpha = 0.9,\beta = 0.1,\gamma = 0.5, \psi = 20$ for all simulations. For global control, we take $\mu = 2.5$ and for local control take $\rho = 7,\delta = 7$. The results for 50\% of the plantation initially infected are shown in Figure \ref{fig:plantation50}, while the results for 10\% of the plantation initially infected are shown in Figure \ref{fig:plantation10}. These results show that under these scenarios the model is not sensitive to the coffee rust initial conditions, and both strategies are effective to control coffee rust. However, in the case of global control, for both sets of initial conditions presented, the coffee rust went to zero, which suggests that in these scenarios global control was faster at reducing coffee rust in comparison to local control.

\subsection{Parameter estimation and uncertainty effects}

In addition to the simulations, we analyzed the estimation of parameters in the model by estimating their probability density function with the Approximate Bayesian Calculation rejection algorithm in a similar way as the one developed in \cite{brown2018}. We analyzed the variables in R using the RNetLogo package to connect NetLogo with R and the EasyABC package to perform the ABC sequencing. We estimated the parameters of both global and local control strategies using the parameter values $\alpha=0.5,\beta=0.8,h=0.07,\gamma=0.3,\psi=1,\mu = 0.6,\rho=0.6,\delta=0.3$. To sample the Approximate Bayesian Calculation, we implemented the Beaumont algorithm using the EasyABC package, based on \cite{beaumont2009}, with a prior uniform distribution on the specific range of each parameter as depicted in Table \ref{param}.

To analyze the uncertainty effect of each parameter, we report the mean, standard deviation, and 95\% confidence interval of each of the accepted samples using the EasyABC package. We used the following statistic as an accepting measure for the sampling process:

\begin{equation}
\Lambda(\hat{H}) = \sqrt{\sum_{t=1}^T\sum_{i=1}^n\sum_{j=1}^m\left(H_{\{i,j\}}(t)-\widehat{H_{\{i,j\}}(t)}\right)^2},    
\end{equation}

\noindent where $\hat{H}$ is the matrix of simulated values of coffee rust with entries $\widehat{H_{\{i,j\}}(t)}$ and $H_{\{i,j\}}(t)$ is the original simulation keeping all the parameters fixed. The expected value of this summary statistic corresponds to the mean of $1000$ simulations with the same parameter values as $H_{\{i,j\}}(t)$ and the tolerance values correspond to $\hat{\sigma}$ and $2\hat{\sigma}$, where $\hat{\sigma}$ corresponds to the standard deviation of $1000$ simulations with the same parameter values as $H_{\{i,j\}}(t)$. The results of these simulations are found in Table \ref{sensitivity}.

\begin{table}[h]
    \begin{center}
    \begin{tabular}{c c | c c c | c c c }
    \hline
        & & \multicolumn{3}{c}{Global} & \multicolumn{3}{c}{Local}\\
        \hline
        Parameter & True Value & Mean & SD & 95\% CI & Mean & SD & 95\% CI\\
        \hline
        $\alpha$ & 0.500 & 0.525 & 0.178 & [0.237,0.947] & 0.463 & 0.227 & [0.112,0.858]\\
        $\beta$ & 0.800 & 0.802 & 0.024 & [0.765,0.839] & 0.831 & 0.077 & [0.699,0.958]\\
        $\gamma$ & 0.300 & 0.298 & 0.012 & [0.280,0.319] & 0.302 & 0.025 & [0.267,0.348]\\
        $h$ & 0.070 & 0.152 & 0.059 & [0.067,0.201] & 0.040 & 0.021 & [0.007,0.076]\\
        $\psi$ & 1.000 & 1.028 & 0.515 & [0.214,2.038] & 1.07 & 0.595 & [0.204,2.426]\\
        $\mu$ & 0.600 & 0.599 & 0.028 & [0.545,0.654] & & &\\
        $\rho$ & 0.600 & & &  & 0.572 & 0.130 & [0.328,0.820]\\
        $\delta$ & 0.300 & & &  & 0.288 & 0.090 & [0.128,0.454]\\
        \hline
    \end{tabular}
    \end{center}
    \caption{Mean, standard deviation, and 95\% confidence interval of the estimate of each parameter using the EasyABC package under global and local control strategies. The parameters $\mu,\rho,\delta$ are only measured in their respective control srategies.}
    \label{sensitivity}
\end{table}

These results let us understand how sensitive the parameters are when they are estimated and also the degree of identifiability of the parameters by analyzing how close the true values are with respect to the center of each corresponding confidence interval. Notice that in most cases the true value lies within a relative distance to this center of less than $10\%$, which tells us that these parameters can be successfully recovered from simulations. This can be appreciated in particular with the $\beta$ and $\gamma$ parameters, which have a low variance relative to their mean, which means that those parameters are not very sensitive to sample variation. This also let us notice that both control strategies are relatively robust, and the estimation process of their parameters with a real data set is feasible.

The only parameter which failed to be recovered is the $h$ parameter, which could be explained by the possibility that for a relatively small value of $h$, the influence of other parameters to the overall behaviour of the system could be more important than the effect of $h$, which could be explained by the high variance relative to the mean values and the relatively large confidence intervals.

\section{Discussion}
We have analyzed a stochastic model that describes the dispersal of coffee rust in a coffee plantation and its interaction with antagonistic bacteria. When introducing bacteria into the system, two strategies were explored, local and global control of coffee rust. With global control, we found at equilibrium that the proportion of infected tree leaves with coffee rust is directly proportional to how much the coffee rust both grows and spreads inside the coffee plantation, and inversely proportional to how much bacteria population is present in each tree. Hence, it is viable to reduce coffee rust in coffee plantations using bacteria, which results in the prevention of problems related to the use of fungicides, such as damage to the soil health \cite{loland2004} and the reduced resistance of coffee trees to other plagues such as the green scale \cite{vandermeerj2014}.

With local control, since the irrigation rate of bacteria is not homogeneous throughout the coffee plantation, when analyzing equilibria of the model, there may be cases where we have different \textit{basic reproductive number}s for each tree. This could lead to the possibility of having $\mathcal{R}_0>1$ in some areas of the plantation, while also $\mathcal{R}_0<1$ in others, which can lead to an unstable behavior near its respective equilibrium point. Albeit an interesting possibility, we presume that additional hypotheses need to be made to explore this scenario.

Furthermore, qualitatively the local and global control strategies do not show a significant difference. In fact, the effect of the irrigation rates of bacteria ($\mu,\rho,\delta$) for the values presented in the framework of Table \ref{param} did not differ greatly, and the results mainly depended in a balance between the coffee rust growth rate ($h$) and the conversion rate of bacteria to coffee rust ($\gamma$). In other words, the coffee rust growth inside a tree depends mostly in its natural growth and the effect the bacteria population has on this population.

For small values of $h$ (coffee rust growth rate), the coffee rust spores decrease rapidly, and increases when the growth rate decreases and the bacteria conversion rate increases. However, when the coffee rust growth rate is high, the coffee rust spores grow steadily even if the bacteria conversion rate is close to 1, therefore the bacteria can reduce the coffee rust of a completely infected leaf by itself. Moreover, the immigration and emigration rates only determined how fast the coffee rust dispersed through the plantation and had little to no effect on the infected leaves in the plantation.

There are several challenges when applying this model in a real-life context. One comes from the environmental dependence of the unknown parameters. Several environmental variables have to be put into consideration, such as amount of rainfall, temperature, and altitude to get an explicit value of the parameters $h,\alpha,\beta$ \cite{nutmanf1963}. Although there is no reported value of the conversion rate of bacteria to coffee rust changing by environmental factors, this possibility cannot be discarded. Another situation comes from calculating the $R_0$ value, which faces the problem of determining the amount of coffee rust in equilibrium if the only factor considered is dispersal ($\Pi$), which depends on plantation size and the probability $P_{\{i,j\}}$. Therefore, these parameters should also be calculated in a case by case basis. However, this model defines a theoretical framework that can be used for developing biological control solutions to coffee producers and whose parameters have been shown to be possible to determine. The equilibrium could be even further analyzed if a method for empirically determining the value of the parameter $\Pi$ is developed.

In conclusion, although we found no significant difference between the effect of global control and local control strategies, there are important differences in the application of each strategy. The global control strategy works like a safety net for the coffee plantation, where most of the plantation is being protected in a simple manner, however its application cost is higher than the application of the local control strategy, especially in bigger plantations, since it would require a larger labor force applying the solution throughout the coffee plantation. 

On the other hand, although the local control strategy is more cost-effective, since it works as a reactive strategy, the plantation has to be in constant monitoring to attack coffee rust as soon as it appears. This requires workers to diligently check the coffee plantation to detect coffee rust in time, which incurs in higher production costs. Although, we can argue these costs are lower compared to a frequent application of the bacteria solution in a global control strategy. Therefore, under this scenario, a global strategy is ideal for smaller plantations, while trying to optimize costs in larger coffee plantations using local strategy.

\section*{Acknowledgements} 
The authors acknowledge the feedback and orientation given by the attendees of the XIX National Meeting of Mathematical Biology at Universidad de Colima, Mexico, and Noel Molina and the agronomists met at the Field Trip of Coffee Producers held in Alajuela, in October 2017. This research did not receive any specific grant from funding agencies in the public, commercial, or not-for-profit sectors.

\medskip

\end{document}